\documentstyle[eqsecnum,prb,aps,twocolumn,epsf]{revtex}

\begin{document}
\draft

\title{Localization and exact current compensation\\
in the quantum Hall effect}
 \author{K. Shizuya}
  \address{Yukawa Institute for Theoretical Physics\\
 Kyoto University,~Kyoto 606-01,~Japan}

\maketitle
 
\begin{abstract} 
A field-theoretic formulation of a planar Hall-electron system with
edges is presented and some fundamental aspects of the integer quantum
Hall effect are studied with emphasis on clarifying general
symmetry-based consequences of localization. 
It is shown, in particular, that the immobility of localized electron
states and current compensation by extended electron states, both
crucial for quantization of the Hall conductance, are
derived through the operation of magnetic translation of localized
electron states alone. They actually are consequences of gauge
invariance and hold under general circumstances with both level mixing
and electron edge states taken into account.
\end{abstract}
 

\section{Introduction}

One of the key elements in the quantum Hall 
effect~\cite{V,Rev,AA,Pr,L,H,TK,LLP,MS,SKM,Wen,KS,KStwo} (QHE) is
localization of electron states via disorder.  
The current-carrying properties of Hall electrons are substantially
modified by localization: The localized electron states cease to
support current. Miraculously, the surviving extended electron states
carry more current and exactly compensate for the loss due to
the localized states. Aoki and Ando~\cite{AA} and Prange~\cite{Pr}
identified this phenomenon of current compensation as a mechanism
responsible for the exact quantization of the Hall conductance;  
it was originally noted in some specific cases (of a strong magnetic
field or a single impurity), but it actually holds under more general
circumstances with level mixing and electron edge states
properly taken into account.~\cite{KStwo}

In view of the principal importance of this phenomenon of current
compensation it is desirable to fully explore its content and
implications. 
The purpose of the present paper is to elaborate on this point.
It was shown earlier~\cite{KS} for an infinite Hall-electron system 
that the current compensation theorem is derived through 
magnetic translation of localized electron states.
In this paper we present a refined version of our previous
consideration adapted to accommodate the presence of sample edges, and
point out that the symmetry principle underlying current compensation 
is electromagnetic gauge invariance.  Special care is taken to treat
inter-level degeneracy inherent to edge states.

In Sec.~II we present a field-theoretic description of 
a planar Hall system with sharp edges.
In Sec.~III we construct the Hamiltonians projected to each
impurity-broadened Landau subband. 
In Sec.~IV we study consequences of localization and give a general
proof of current compensation.
In Sec.~V we present another proof within a linear-response treatment,
which shows an interplay of the localized and extended states
explicitly.  Section~VI is devoted to concluding remarks.\\

\setcounter{section}{1}
 \section{Field theory of Hall electrons}

In this section we formulate a field theory of Hall electrons 
in the presence of disorder. 
Consider electrons confined to an infinitely long strip of width
$L_{y}$ (or formally, a strip bent into a loop of circumference
$L_{x} \gg  L_{y}$), described by the Hamiltonian:
\begin{eqnarray}
&&{\cal H} =\int dx dy \, \Psi^{\dag}(x,y,t)\, {\sf H}\, 
\Psi(x,y,t),   \label{HFT}\\
&&{\sf H} =H_{0}+ U(x,y)-eA_{0}(y), \label{HAU}\\
&&H_{0} = {1\over 2}\omega \Bigl\{ \ell^{2}p_{y}^{2}
+(1/\ell^{2})(y - y_{0} -e\ell^{2}A_{x})^{2}\Bigr\},
 \label{hzero}
\end{eqnarray}
written in terms of $\omega \equiv eB/m$,
the magnetic length ${\ell}\equiv 1/\sqrt{eB}$ and 
$y_{0} \equiv p_{x}/(eB)$. 
Here $U(x,y)$ stands for a random impurity potential and 
the Landau-gauge vector potential $(-By,0)$ has been used
to supply a uniform magnetic field $B$
normal to the plane.
We take explicit account of the two edges 
$y=\pm L_{y}/2$, where the wave function is bound to vanish.

We shall study the Hall effect in a static setting:
The scalar potential $A_{0}(y)$ supplies a general 
Hall field 
$E_{y}(y)= -\partial_{y} A_{0}(y)$ that can vary 
across the strip.
We use a constant potential $A_{x}$ to detect the total 
Hall current $J_{x}=\int dy\, j_{x}(x,y,t)$, or more
precisely, its spatial average $(1/L_{x})\int dx\, J_{x}$,
that flows in response to $E_{y}(y)$.

The kinetic term $H_{0}$ describes the cyclotron motion of 
an electron.
The eigenstates of $H_{0}$ in the sample bulk are 
Landau levels with discrete energy $\omega (n +{1\over{2}})$, 
labeled by integers $n = 0,1,2,\cdots$, and $y_{0}=\ell^{2} p_{x}$. 
The eigenfunctions in the presence of sharp edges are still 
labeled by $N=(n,\, y_{0})$: 
\begin{equation}
\psi_{N}(x,y)\equiv \langle x,y |N\rangle = (2\pi
\ell^{2})^{-1/2}\,e^{ixp_{x}}\phi_{N}(y), \label{psiN}
\end{equation}
with $\phi_{N}(y)$ given~\cite{MS} by the parabolic cylinder 
functions~\cite{WW}  $D_{\nu}(\pm \sqrt{2}(y-\bar{y}_{0})/\ell)$ 
for electrons residing near the edges $y = \mp L_{y}/2$, where
$\bar{y}_{0}\equiv y_{0}+e\ell^{2} A_{x}$.
Setting $\phi_{N}=0$ at $y=\pm L_{y}/2$ fixes the energy eigenvalues
\begin{equation}
\epsilon_{N}= \omega \{{\nu_{n}}(\bar{y}_{0})+1/2 \}
\label{eN} 
\end{equation}
as a function of $\bar{y}_{0}$ for each $n$.
The spectra $\nu_{n}(y_{0})$ are determined numerically.
They rise sharply for  $|y_{0}| \gtrsim L_{y}/2$, and recover the
integer values $n$ as $y_{0}$ moves to the interior a few magnetic
lengths away from $y_{0}= \pm L_{y}/2$;~\cite{MS}
for the $n=0$ level, $\nu_{0}(y_{0})$ decreases from 1 to 0.003 as
$y_{0}$ moves inward from $\pm L_{y}/2$  by $2.5 \ell$.

The normalized wave functions
\begin{equation}
 \phi_{N}(y)= \phi_{n} (y-\bar{y}_{0}; \bar{y}_{0}) 
 \equiv \phi_{n} (y;\bar{y}_{0}) ,
\label{phiN} \end{equation}
taken to be real, depend on $A_{x}$ only through 
$\bar{y}_{0}=  y_{0}+e\ell^{2} A_{x}$. 
They are highly localized around $y\sim \bar{y}_{0}$ with spread
$\triangle y \sim O(\ell)$, and a single-particle state $(n,y_{0})$
has its center-of-mass at position
$y^{\rm cm}= \bar{y}_{0}-\ell^{2}{\nu_{n}}'(\bar{y}_{0})$
on the $y$ axis; see Eq.~(\ref{ynn}) below.
This $y^{\rm cm}$ stays within the sample width $[-L_{y}/2,L_{y}/2]$.
In principle, $y_{0}=\ell^{2} p_{x}$ has an infinite range  
$-\infty < y_{0} < \infty$, but under circumstances of practical
interest (where the number of filled levels is limited)
it falls in a reduced range  
$-L_{y}/2 \lesssim y_{0} \lesssim L_{y}/2$.

For the description of drift motion of an orbiting electron 
it is advantageous to express the electron field $\Psi$ 
in terms of the eigenmodes of $H_{0}$.
Note that in the basis $|N \rangle =|n,y_{0} \rangle$ 
the coordinate $x$ has the representation
\begin{eqnarray}
\langle N| x |N' \rangle
&=& -\ell^{2} [ \delta_{n n'}k  + 
\langle k \rangle_{n n'} ]\, \delta (y_{0}-y_{0}') , 
\label{NxN} 
\end{eqnarray}
with $k=-i\partial/\partial y_{0}$;
likewise, $ \langle N| y |N' \rangle = 
\langle y \rangle_{n n'}\delta (y_{0}-y_{0}')$.
[To obtain Eq.~(\ref{NxN}) replace $x$ by a derivative $-\ell^{2}k$
acting on $e^{-iy_{0} x/\ell^{2}}$ in $\langle N|x,y\rangle$.]
Here we have introduced the notation 
\begin{equation}
 \langle {\cal O} \rangle_{m n} =\int_{-L_{y}/2}^{L_{y}/2} dy\, 
\phi_{m} (y;\bar{y}_{0})\, {\cal O} \,\phi_{n} (y;\bar{y}_{0})     
\label{matrixnotation} 
\end{equation}
with normalization $\langle 1 \rangle_{m n}=\delta_{m n}$.

For later convenience let us denote 
$Y_{mn}(\bar{y}_{0}) \equiv$ 
\mbox{$\langle y - \bar{y}_{0} \rangle_{mn}/\ell$} and
$Q_{mn}(\bar{y}_{0}) \equiv -\ell \langle k \rangle_{mn}$;
$Y_{mn}$ are symmetric in $(m,n)$ while
$Q_{mn}$ are antisymmetric.
They are completely known through the spectra
$\nu_{n}(\bar{y}_{0})$; see Appendix A.
Here we quote only the following: 
\begin{eqnarray}
Y_{n n} &=& \langle y -\bar{y}_{0} \rangle_{n n}/\ell=
-\ell\, {\nu_{n}}'(\bar{y}_{0}),  \label{ynn} \\
Y_{m n}&=& \sigma_{m n}\, \ell\,  |{\nu_{m}}'{\nu_{n}}'|^{1/2}
 /[(\nu_{m}-\nu_{n})^{2}-1],  \label{ymn}\\
Q_{m n}&=& i\, Y_{mn}/(\nu_{m}-\nu_{n})\ \ \ \ (m\not=n), 
\label{qmn}
\end{eqnarray}
where $\nu_{n}=\nu_{n}(\bar{y}_{0})$ and 
${\nu_{n}}'\equiv \partial_{y_{0}}{\nu}_{n}(\bar{y}_{0})$;
the overall sign $\sigma_{m n}=1$ for $y_{0} \sim +L_{y}/2$
while  $\sigma_{m n}=(-1)^{1+n+m}$ for $y_{0} \sim -L_{y}/2$.
The special case $\nu_{m}-\nu_{n}=\pm 1$
arises only in the sample interior  
$|y_{0}|< L_{y}/2 - O(\ell)$, where $Y_{mn}$ and $Q_{mn}$
are reduced to constant hermitian matrices given by 
\begin{equation}
 Y_{mn}^{(\rm bulk)} + i\, Q_{mn}^{(\rm bulk)} =\sqrt{2n}\,
\delta_{m,n-1} .  
\label{yplusip} 
\end{equation}

With the expansion of the electron field 
$\Psi (x,y,t)=\sum_{N}\! \psi_{N}(x,y)\, a_{N}(t)$,
the Hamiltonian~(\ref{HFT}) is rewritten as
\begin{equation}
 {\cal H}=\int dy_{0} \ \sum_{m,n=0}^{\infty}
a^{\dag}_{m}(y_{0},t)\, H_{m n}\, a_{n}(y_{0},t),  
\label{aHa}
\end{equation}
where we have set $a_{N}(t) \rightarrow a_{n}(y_{0},t)$.
The matrix Hamiltonian $H_{m n}$ is given by
\begin{eqnarray}
H_{m n}
&=&  \delta_{m n}\,\omega \{ \nu_{n}(\bar{y}_{0})\!+\!1/2\}
\nonumber\\
& &+ [\bar{U}(\bar{y}_{0},k)]_{m n} 
-e [A_{0}(\bar{y}_{0}\! +\!\ell\, Y)]_{m n} .
\label{Hmn}
\end{eqnarray}
where 
\begin{equation}
\bar{U}(\bar{y}_{0},k)\equiv 
U(-\ell^{2}k\! +\!\ell Q,\bar{y}_{0}\!+\!\ell Y )
\label{Ubar}
\end{equation}
with $k\equiv -i\partial /\partial y_{0}$ 
stands for the impurity potential in the
$N$ representation~\cite{fn} obtained from $U(x,y)$ through
substitution $x \rightarrow\! -\ell^{2}k +\ell\, Q$ and 
$y \rightarrow  \bar{y}_{0} +\ell\,Y$.
[From now on we shall frequently suppress Landau-level 
indices and employ matrix notation; notation $[\cdots]_{mn}$ refers to
some specific components.]

In the Hamiltonian~(\ref{aHa}) the motion of a Hall electron is
decomposed into relative cyclotron motion described by matrix dynamics
and c.m.~motion described by a one-dimensional field theory with
coordinate $y_{0}$ and its conjugate $k$. 
Physically $x_{0}\!=\! -\ell^{2}k$ and $\bar{y}_{0}$ stand for the
center coordinates of an orbiting electron. They are the generators of
magnetic translation,~\cite{Zak} and obey 
$[x_{0},\bar{y}_{0}]=i\ell^{2}$.
Note that the relative coordinates $Y$ and $Q$ obey a nontrivial
commutation relation  
\begin{equation} 
[Y, Q] = i\,(1 + \ell\,\partial_{y_{0}}Y ),  
\label{YQcom} 
\end{equation}
which follows from the trivial relation $[x,y]=0$. 

The presence of sample edges has effectively generated strong
potentials $\omega \nu_{n}(\bar{y}_{0})$ that confine Hall electrons
to finite width $-L_{y}/2 \lesssim y_{0} \lesssim L_{y}/2$ 
in $y_{0}$ space.  These confining potentials drive orbiting electrons
and thus make electrons residing near the sample edges 
characteristically different from those in the sample bulk.
Indeed, as seen clearly in the impurity-free $(U\rightarrow0)$ case
where an electron state $(n,y_{0})$ acquires the group velocity 
\begin{equation}
v_{x}= \omega \ell^{2} {\nu_{n}}'(\bar{y}_{0}) +
        E_{y}(\bar{y}_{0})/B +\ldots,
	\label{vgroup}
\end{equation}
near the sample edges the electrons travel with velocity 
$v_{x}\approx \omega \ell^{2} {\nu_{n}}'
\sim \pm \omega \ell$
much larger than the field-induced drift velocity $E_{y}/B$.
[Numerically $v_{x}^{\rm edge}\sim \omega \ell \sim 10^{7}$~cm/s for
typical values $\omega \sim 10~$meV and $\ell \sim$ 100~{\AA} while 
$v_{x}^{\rm bulk}\sim E_{y}/B \sim 10^{3}$~cm/s for 
\mbox{$E_{y}$ = 1~V/cm} and $B$ = 5~T.]
Classically these edge states~\cite{H,MS,SKM,Wen} are
visualized as electrons hopping along the sample edge.~\cite{P}
They travel in opposite directions at opposite edges, and the currents
they carry at the two edges combine to cancel in equilibrium.~\cite{H}
For distinction we refer to electron states in the sample bulk as
bulk states.

\section{subband Hamiltonians}

In this section we derive Hamiltonians projected to each
impurity-broadened Landau level. To start with let us note that
a concise expression for the $x-$averaged current of our interest, 
$J_{x}=(1/L_{x}) \int dx \int dy j_{x}$, is 
\begin{equation}
J_{x}= - (1/L_{x}) \int dxdy\, \Psi^{\dag}\, 
\partial_{A} {\sf H}\,\Psi,  
\label{Jxx}
\end{equation}
where $\partial_{A} {\sf H} \equiv \partial {\sf H}/\partial A_{x}$.
The use of this representation lies in the fact that 
the relevant current is calculated from the $A_{x}$ dependence 
of the eigenvalues of the Hamiltonian ${\sf H}$.

Our discussion below makes extensive use of unitary transformations.
It will therefore be useful to consider how a unitary transformation
$G$, which may in general depend on $A_{x}$, affects the current
operator~(\ref{Jxx}).  With the transformation 
$\Psi \rightarrow \Psi^{G}= G\, \Psi$ and 
${\sf H} ^{G} = G {\sf H} G^{-1}$, $J_{x}$ reads
\begin{equation}
J_{x}= - (1/L_{x}) \int dxdy\, (\Psi^{G})^{\dag}\, \left\{
\partial_{A} {\sf H}^{G} - 
[\partial_{A}G\,G^{-1},\, {\sf H}^{G}] \right\}\Psi^{G},  
\label{Jxxtr}
\end{equation}
Note that the commutator term has a vanishing expectation value 
$\langle \alpha| [\partial_{A}G\,G^{-1}, {\sf H}^{G}] |\alpha\rangle
=0$
for each eigenstate $|\alpha\rangle$ of ${\sf H}^{G}$ as long as 
the matrix element 
$\langle \alpha | \partial_{A}G\,G^{-1} |\alpha\rangle$  
exists. Thus, when $G$ itself is a well-defined operator, 
no explicit account of the commutator term is needed in 
calculating the physical expectation value of the current $J_{x}$, 
which is determined solely from the $\partial_{A} {\sf H}^{G}$ part.

As an example, the wave functions $\langle x,y |N \rangle$
play the role of a unitary transformation 
$G_{N {\bf x}}= \langle N |{\bf x} \rangle$ that connects 
the ${\bf x}=(x,y)$ and $N=(n,y_{0})$ representations.  
The $A_{x}$ dependence of this $G$ is characterized by $Q$:
\begin{equation}
 \langle N|\partial_{A}G\,G^{-1}|N' \rangle =
ie\ell\, \delta (y_{0}-y_{0}')\, Q_{n n'}(\bar{y}_{0}).
 \label{dGG}
\end{equation} 
The current $J_{x}$, written as $J_{x} \propto \int dy_{0}
a^{\dag}_{m} Y_{mn} a_{n}$ in the $N$ representation, 
therefore differs from an equivalent representation 
$J_{x} \propto \int dy_{0} a^{\dag}_{m} (\partial_{A} H_{mn}) a_{n}$
by a term $\propto [Q,H]$.

Let us take local disorder into account. Impurities scatter
electrons and turn each Landau level into a broadened subband. 
When the disorder $U(x,y)$ is weak compared with the level gap 
$\sim \omega$, one can diagonalize the Hamiltonian~(\ref{Hmn}) 
with respect to Landau-level labels by a unitary transformation 
of the form $[W(\bar{y}_{0},k)]_{mn}$:
\begin{equation}
H^{\rm sb}(\bar{y}_{0},k)= W\,\bigl\{ \omega\,
\nu(\bar{y}_{0}) 
+\omega/2 + \hat{U}\! -e A_{0} \bigr\}\,W^{-1}, 
\label{Hsb}
\end{equation}
where $ \nu(\bar{y}_{0})={\rm diag}[ \nu_{n}(\bar{y}_{0})]$.
In particular, when the edge states are ignored, 
one may adjust the operator-valued matrix $[W(\bar{y}_{0},k)]_{mn}$ 
successively to each power of $(\bar{U}\!-\!e A_{0})/\omega$ so that
off-diagonal terms disappear from $H^{\rm sb}$, obtaining a subband
Hamiltonian $[H^{\rm sb}]_{nn}$ of the form
\begin{eqnarray}
&& \omega\,(n + 1/2) + U^{A}_{nn}(\bar{y}_{0},k) \nonumber \\
&&\ \ +{1\over{\omega}} \sum_{m\not=n}{1\over{n-m}}\,
U^{A}_{nm}(\bar{y}_{0},k)\,U^{A}_{mn}(\bar{y}_{0},k) + \cdots,
\end{eqnarray}
where $U^{A}_{mn}\equiv 
[\bar{U}(\bar{y}_{0},k) - e A_{0}(\bar{y}_{0}\!+\!\ell\,Y)]_{mn}$.

Edge states are afflicted with inter-level degeneracy; i.e.,
the edge states of a given level get degenerate in energy with
some electron states of the lower levels.
Still it is possible to carry out diagonalization by a unitary 
transformation of the form $W(\bar{y}_{0},k)$.
See Appendix B for details. 
Actually, relevant to our discussion below is only the fact 
that $W$ and $H^{\rm sb}$ consist of $\bar{y}_{0}$ and $k$.

With level mixing now properly taken care of, 
each subband becomes independent:
\begin{equation} 
{\cal H}=\int dy_{0} \ \sum_{n=0}^{\infty}
b^{\dag}_{n}(y_{0},t)\, [H^{\rm sb}(\bar{y}_{0},k)]_{n n}\,
b_{n}(y_{0},t), 
\label{Htr}
\end{equation}
where $b_{n}(y_{0},t)=\sum_{m}[W(\bar{y}_{0},k)]_{nm}a_{m}(y_{0},t)$.
Correspondingly we shall henceforth concentrate on a single subband, 
say, the $n$th one, and write 
\begin{equation}
\hat{H}(\bar{y}_{0},k)\equiv [H^{\rm sb}(\bar{y}_{0},k)]_{nn},
\label{Hsbnn}
\end{equation}
for short.
Let $\{ |\alpha\rangle \}$ denote the eigenstates of
$\hat{H}(\bar{y}_{0},k)$,  which are taken to form an orthonormal set.
The current $J_{x}$ the $n$th subband supports is calculated from 
the $A_{x}$ dependence of the spectrum:
\begin{equation}
J_{x}^{\{n\}}= -{e\ell^{2}\over{L_{x}}} \sum_{\alpha}
 \langle \alpha|\partial_{y_{0}}\hat{H}(\bar{y}_{0},k) 
|\alpha\rangle,
\label{Jxnexact}
\end{equation}
where the sum $\sum_{\alpha}$ runs over all the occupied states
$\alpha$ within the $n$th subband.

The Hamiltonian $\hat{H}(\bar{y}_{0},k)$ as an operator formally acts
on the $n$th subband spanned by the $y_{0}-$diagonal basis 
$\{ |y_{0}\rangle ; -\infty < y_{0} < \infty\}$, which forms 
a complete set of the eigenstates of $\hat{H}(\bar{y}_{0},k)$
with $k\rightarrow 0$. 
(For brevity suffix $n$ for $|y_{0}\rangle$ is suppressed.)
Here associated with the basis vectors $|y_{0}\rangle$ are 
the coordinate-space wave functions 
$\langle x,y|W^{-1}|n,y_{0}\rangle$,
which, owing to dominant admixture of the plane-wave mode 
$\langle x,y|n,y_{0}\rangle$, are extended in $x$.

\section{ Localization and its consequences}

In the absence of impurities all electron states
$|n,y_{0}\rangle$ are plane waves extended in $x$, 
though localized in $y$ with spread $\triangle y \sim O(\ell)$.
Such spatial characteristics of electron states are modified in the
presence of disorder: Electrons, scattered repeatedly by
impurities, tend to be confined in finite domains of space, and
in two-dimensional Hall samples the majority of electron 
states are considered to be localized.~\cite{AA,Pr,L}
In this section we study consequences of localization by use of
magnetic translation.

\subsection{Localized states v.s.~extended states}

As preliminaries, note first that the subband Hamiltonian 
$\hat{H}(\bar{y}_{0},k)$ depends on $A_{x}$ only through  
$\bar{y}_{0}=  y_{0} + e\ell^{2}\,A_{x}$, 
and try to translate each electron mode by $-e\ell^{2}A_{x}$
in the $y_{0}$ direction:
\begin{equation}
b_{n}^{\rm trans}(y_{0},t) = e^{-ie \ell^{2} A_{x} k} b_{n}(y_{0},t),
 \label{transl}
\end{equation}
where $k=-i\partial/\partial y_{0}$. 
This makes the transformed Hamiltonian 
$e^{-ie \ell^{2} A_{x} k} \hat{H}(\bar{y}_{0},k) 
e^{ie \ell^{2} A_{x} k} = \hat{H}(y_{0},k)$ independent of $A_{x}$.
Thus this transformation, if allowed to perform, would imply that
there is no current. A resolution to this apparent puzzle turns out
revealing, as explained below.

In connection with $J_{x}$ in Eq.~(\ref{Jxxtr}) 
we have noted that a commutator of the form 
$[{\cal O},{\sf H}^{G}]$ has a vanishing expectation value 
for each eigenstate $|\alpha \rangle$ of ${\sf H}^{G}$.
Such expectation values need not vanish in case 
$\langle \alpha |{\cal O} |\alpha \rangle$ are ill-defined or
singular. For example, consider the impurity-free 
Hamiltonian $h_{0}= \omega \{ \nu (\bar{y}_{0}) + 1/2\}$ and 
its eigenstates $|N\rangle$.
While the commutator $i[k, h_{0}] 
= \omega  \nu_{n}'(\bar{y}_{0})$ itself is a well-defined operator,
its expectation values 
$\langle N |[i k, h_{0}] |N \rangle /\langle N | N \rangle 
= \omega \nu_{n}'(\bar{y}_{0})$ 
fail to vanish because 
$\langle N | k |N \rangle /\langle N | N \rangle
\sim O(L_{x}/\ell^{2})$ are singular. Physically this singular
behavior derives from the fact that the states $|n,y_{0}\rangle$
stand for plane waves with uncertainty 
$\triangle y_{0}\sim 2\pi\ell^{2}/L_{x}$ so that 
their $x$ positions are indeterminate, $\triangle x \sim L_{x}$,
in accordance with the uncertainty relation 
$[x_{0},y_{0}]=i\ell^{2}$.  In contrast, the relative coordinates 
$Y_{mn}(\bar{y}_{0})$ and $Q_{mn}(\bar{y}_{0})$ 
are bounded operators of magnitude of $O(1)$, so that 
$\langle N|[Y,h_{0}]|N\rangle = \langle N|[Q, h_{0}]|N\rangle =0$ 
holds trivially.

From this consideration emerges the following characterization of 
localized electron states.
The localized states are states whose wave functions have finite
spatial spread $\triangle x \ll L_{x}$ and $\triangle y \ll L_{y}$.  
Actually, for the finite-width system of our present concern where 
we detect the current flowing in the $x$ direction, 
it does not matter whether electron states are localized in $y$ or 
not.  Correspondingly, we here call localized states 
only those that have finite spatial spread $\triangle x$ in $x$. 
That is, the notion of the c.m.~position $x_{0}= -\ell^{2}\,k$ 
(in the $x$ direction) has a definite meaning for localized states.
Let us rephrase this in more definite terms: 
For each localized state $|\lambda \rangle$ the state $k |\lambda 
\rangle$ exists; i.e., matrix elements 
like $\langle \lambda' | k |\lambda \rangle$ 
and $\langle \sigma | k |\lambda \rangle$ are well-defined.
[For ease of notation we shall henceforth reserve $\lambda, 
\lambda',\cdots$ for localized states, $\sigma,
\sigma',\cdots$ for extended states, and $\alpha, \beta$ 
for general states.]
For example, a localized state $|\lambda)$ (of the $n=0$ level) 
due to a short-range impurity located at $x=y=0$ has 
a wave function of the form $\langle y_{0}|\lambda)\propto
e^{-{1\over2}y_{0}^{2}/\ell^{2}}$, for which
$\langle y_{0}|k|\lambda)\propto i (y_{0}/\ell^{2}) 
e^{-{1\over2}y_{0}^{2}/\ell^{2}}$ is well-defined. 
In contrast, for extended states analogous states $k |\sigma\rangle$
are ill-defined; in particular, 
$\langle \sigma| k | \sigma\rangle \sim O(L_{x}/\ell^{2})$ are
singular.

The solution to the puzzle is now clear: The magnetic translation 
$e^{-ie \ell^{2} A_{x} k}$ is ill-defined for extended states and,
in particular, the commutator term $[k,\hat{H}]$ has introduced a
substantial modification of the current operator. The current $J_{x}$
thus can no longer be derived from the $A_{x}$ dependence of the
transformed Hamiltonian.

\subsection{Immobility of localized states}

Since the transformation makes sense for localized states, one may
still choose to translate them and ask what would happen. 
An immediate result is that the localized states cease to support
current, as shown below.

For the proof we shall examine the $A_{x}$ dependence of the exact
spectrum of $\hat{H}(\bar{y}_{0},k)$ by starting with the zeroth-order
Hamiltonian $\hat{H}^{(0)}(y_{0},k)\equiv \hat{H}(y_{0},k)$.
Let $\{ |\alpha) \}$ denote the eigenstates of $\hat{H}^{(0)}$, 
which are obtained from the eigenstates
$\{ |\alpha\rangle \}$ of $\hat{H}$ by letting $A_{x} \rightarrow 0$.
We divide them into localized states $\{\, |\lambda)\,\}$
and extended states $\{\, |\sigma)\,\}$, i.e., 
$\hat{H}^{(0)}(y_{0},k)\,\langle y_{0}|\alpha)
= \epsilon_{\alpha} \langle y_{0}|\alpha)$ for
$\alpha=\lambda, \sigma$; obviously both eigenvalues
$\epsilon_{\alpha}$ and eigenfunctions $\langle y_{0}|\alpha)$
are independent of $A_{x}$.
We suppose that $A_{x}$ is very weak so that both $|\alpha)$ and 
the paired state $|\alpha\rangle$ share the same extended or 
localized character.

With the expansion of the field operators 
$b_{n}(y_{0},t)= \sum_{\alpha} \langle y_{0}|\alpha) b_{\alpha}(t)$
in terms of the eigenfunctions of $\hat{H}^{(0)}$,
the Hamiltonian (\ref{Htr}) for the $n$th subband is rewritten as   
${\cal H}_{n}\equiv \sum_{\alpha, \beta} b_{\alpha}^{\dag}\,
\hat{H}_{\alpha\beta} b_{\beta}$,
where $\hat{H}_{\alpha\beta} \equiv 
(\alpha|\hat{H}(\bar{y}_{0},k)|\beta)$.
We now translate only the localized modes $b_{\lambda}$
so that
\begin{equation}
c_{\lambda} = T_{\lambda\lambda'} b_{\lambda'} \ \ \ \ 
{\rm with}\ \  T= e^{- i e\ell^{2}A_{x}k}, 
\label{Tb}
\end{equation}
where $T_{\lambda\lambda'}\equiv (\lambda|T|\lambda')$.
Then the Hamiltonian reads
\begin{eqnarray}
{\cal H}_{n}= &c_{\lambda}^{\dag}& T_{\lambda\lambda''}
\hat{H}_{\lambda''\lambda'''}
(T^{-1})_{\lambda'''\lambda'}\, c_{\lambda'}  
+b_{\sigma}^{\dag}\, \hat{H}_{\sigma\sigma'}\, b_{\sigma'} \nonumber\\
&+& c_{\lambda}^{\dag}\, T_{\lambda\lambda'}\, 
\hat{H}_{\lambda'\sigma}\, b_{\sigma}
+ b_{\sigma}^{\dag}\,\hat{H}_{\sigma\lambda'}\, 
(T^{-1})_{\lambda'\lambda} c_{\lambda},  
\label{Hn}
\end{eqnarray}
where $\hat{H}_{\lambda\lambda'} \equiv 
(\lambda|\hat{H}(\bar{y}_{0},k)|\lambda')$, etc.
Summations over repeated labels $\lambda, \sigma, ...$~are understood
from now on.

The $T_{\lambda\lambda'}$ is a unitary transformation within the space
of localized states $\{|\lambda)\}$.  Actually, in our discussion
below it suffices to fix $T$ only to $O(A_{x})$ and thus to take 
$T_{\lambda\lambda'}= \delta_{\lambda\lambda'} 
-  i e\ell^{2}A_{x} (\lambda|k|\lambda') + \cdots$.

Let us next rewrite the $T \hat{H} T^{-1}$ term in Eq.~(\ref{Hn}) 
in terms of the $y_{0}$-diagonal basis 
$\{\, |y_{0}\rangle\,\}= \{\,|\lambda)\,\} + \{\,|\sigma)\,\}$: 
\begin{eqnarray}
&&T_{\lambda\lambda''}\hat{H}_{\lambda''\lambda'''}
(T^{-1})_{\lambda'''\lambda'}      
= T_{\lambda y_{0}}\,\hat{H}_{y_{0} y_{0}'} (T^{-1})_{y_{0}'\lambda'} 
+ {\cal C}_{\lambda\lambda'}, \nonumber  \\
\label{tht} \\
&&{\cal C}_{\lambda\lambda'}=
-T_{\lambda\sigma}\hat{H}_{\sigma\lambda''}(T^{-1})_{\lambda''\lambda'} 
-T_{\lambda\lambda''}\hat{H}_{\lambda''\sigma}
(T^{-1})_{\sigma\lambda'} \nonumber  \\
&&\hskip1.2cm 
-T_{\lambda\sigma}\hat{H}_{\sigma\sigma'}(T^{-1})_{\sigma'\lambda'} ,
\label{Olambda} 
\end{eqnarray}
where $T_{\lambda y_{0}}\equiv (\lambda|T|y_{0}\rangle$, 
$\hat{H}_{y_{0} y_{0}'}\equiv \langle y_{0}|\hat{H}|y_{0}'\rangle$, etc.
Here, through the introduction of the transformation 
\begin{equation}
T_{\lambda y_{0}}=(\lambda|T|y_{0}\rangle =
(\lambda|y_{0}\rangle  - ie\ell^{2} A_{x} (\lambda|k|y_{0}\rangle 
+\cdots,
\label{Tly}
\end{equation}
we have let $T$ act on the extended states $\{\,|\sigma)\,\}$ as well.
Note that the transformations involved in Eq.~(\ref{tht}) such as
$T_{\lambda \sigma}= -ie\ell^{2} A_{x} (\lambda|k|\sigma) +\cdots$ 
are all well-defined to $O(A_{x})$;
they do not involve any matrix elements of the kind
$(\sigma|k|\sigma')$ connecting extended states.

As desired, the $T_{\lambda y_{0}}\,\hat{H}_{y_{0} y_{0}'} 
(T^{-1})_{y_{0}'\lambda'}$ term in Eq.~(\ref{tht}) becomes
independent of $A_{x}$:
\begin{equation}
(\lambda|T \hat{H}(\bar{y}_{0},k) T^{-1}|\lambda') 
= (\lambda|\hat{H}(y_{0},k)|\lambda'), 
\label{thttwo}
\end{equation}
which equals $\epsilon_{\lambda} \delta_{\lambda \lambda'}$. 
By virtue of $T_{\lambda\sigma}=O(A_{x})$ and 
$\hat{H}_{\sigma\lambda}=O(A_{x})$, 
the remaining term ${\cal C}_{\lambda\lambda'}$ is only
of $O(A_{x}^{2})$ and no other terms in ${\cal H}_{n}$ of
Eq.~(\ref{Hn}) give rise to an $O(A_{x})$ energy shift for 
each localized state $\lambda$.
This proves that the localized states
support no current, 
i.e., $\langle \lambda|\partial_{y_{0}}\hat{H}(\bar{y}_{0},k) 
|\lambda\rangle =0$.

\subsection{Edge states v.s.~bulk states}

Nontrivial consequences of disorder all derive from the $x$ dependence
of the impurity potential $U(x,y)$ or the $k$ dependence of
$\bar{U}(\bar{y}_{0},k)$.   
Indeed, if one sets $k\rightarrow 0$ in $\hat{U}$, the Hamiltonian
$\hat{H}(\bar{y}_{0},k)$ becomes local in $\bar{y}_{0}$,
and all the eigenstates are plane waves 
extended in the $x$ direction, leading to no localization.

Physically this $k$ dependence is tied to the 
impurity-induced drift of each electron in the $y$ direction with
velocity $v_{y}\sim \ell^{2}\partial_{x}U$.
Electrons in the sample bulk, drifting slowly with velocity 
$v_{x}^{\rm bulk} \sim E_{y}/B$, are readily scattered by impurities. 
In contrast, the electrons at the sample edges, 
traveling faster with velocity $v_{x}^{\rm edge} \sim \omega \ell $, 
are much less sensitive to such impurity-induced drift.
Accordingly the spread $\triangle y_{0}$ of the wave functions
$\langle y_{0}|\alpha)$ in $y_{0}$ space provides a natural measure 
to distinguish between the edge and bulk states:
The bulk states acquire typical spread $\triangle y_{0} \gtrsim \ell$ 
whereas the edge states have smaller spread 
$\ell^{2}/L_{x}\lesssim \triangle y_{0} < \ell$.
(Remember here that a plane-wave state $|y_{0}\rangle$ has tiny spread 
$\triangle y_{0}= 2\pi \ell^{2}/L_{x}$ in $y_{0}$ space while it has 
spread $\triangle y \sim O(\ell)$ in real space.)
The spread $\triangle y_{0}$ gets even smaller 
as an edge state travels faster, i.e., as it gets closer to the edge;
this is readily verified by a numerical simulation.

The edge states are thus close to plane-wave states, though not 
identical, and this feature provides a way to label them.
Suppose that, as we let $k \rightarrow 0$ in $\hat{H}(\bar{y}_{0},k)$,
an edge state $|\alpha\rangle$ is reduced to a plane-wave state 
$|y^{\rm e}_{0}\rangle$ of $\hat{H}(\bar{y}_{0},0)$.  We can use 
$y^{\rm e}_{0}$ to label the edge states $|\alpha\rangle$, i.e., set 
$\alpha \rightarrow y^{\rm e}_{0}$ and 
$\sum_{\alpha} \propto \int d y^{\rm e}_{0}$. 
Then the edge-state spectrum 
$\epsilon_{\alpha}=\epsilon(\bar{y}^{\rm e}_{0})$ 
depends on $A_{x}$ only through $\bar{y}^{\rm e}_{0}=y^{\rm e}_{0} 
+e\ell^{2}A_{x}$.
[Note here that physical quantities depend on $p_{x}$ only through 
the combination of a covariant derivative 
$\bar{y}_{0}= \ell^{2} (p_{x} + e A_{x})$; 
this $\bar{y}^{\rm e}_{0}$ dependence of the edge-state spectrum 
is also directly verified by calculating $\epsilon_{\alpha}$
perturbatively from the spectrum of $\hat{H}(\bar{y}_{0},0)$.] 
We suppose that the edge-state spectrum $\epsilon(\bar{y}^{\rm e}_{0})$ 
continuously rises (or falls) with $y^{\rm e}_{0}$, as in the
impurity-free case. For extremely fast electrons the influence of
disorder can be neglected so that  
$\epsilon(\bar{y}^{\rm e}_{0}) \rightarrow 
\hat{H}(\bar{y}^{\rm e}_{0},0)$ 
for $ y^{\rm e}_{0} \rightarrow \pm \infty$.

In general there is such one-to-one correspondence between the
eigenstates of $\hat{H}(\bar{y}_{0},k)$ and those of
$\hat{H}(\bar{y}_{0},0)$ (because the electron number is
conserved). Unlike the edge-state spectrum, however, the bulk-state
spectrum is not a smooth function of $y^{\rm e}_{0}$.

\subsection{Quantum Hall effect}

We are now ready to study the quantum Hall effect.
Suppose we fill a subband with electrons by varying $B$ or electron 
population gradually.
Since the confining potential is higher near the sample edges,
the electrons are accommodated in the sample bulk first.
No net current flows while only localized states arise, and a lower
Hall plateau develops.
The current starts to flow as soon as extended states emerge. 
When the extended states in the bulk-state spectrum are filled up, 
the edge states begin to emerge. 
Figure~1 schematically shows the spectrum of one such well-filled subband 
when a Hall field is turned on. [The spectrum gets tilted in the
presence of a potential $A_{0}(y)$.]
There each electron state is labeled in terms of the related
eigenstate of $\hat{H}(\bar{y}_{0},0)$; the $y_{0}$ axis therefore
refers to $y_{0}^{\rm e}$.
The bulk-state spectrum, being discontinuous in $y_{0}^{\rm e}$, is
depicted as a broadened spectrum lying in the interval 
$\eta^{-}\le y_{0}\le \eta^{+}$. 
The edge-state spectrum $\epsilon(\bar{y}^{\rm e}_{0})$ is 
drawn with solid curves, and is filled up to the energy 
$\epsilon^{\pm} \equiv \epsilon(y^{\pm}_{0})$ at the two sample edges.

Suppose now that the extended states in the bulk-state spectrum
are filled up in Fig.~1.
Then this subband supports a fixed amount of current determined 
by the Hall voltage alone, as explained below.
It is clear that one can regard the localized states as all occupied
in calculating the current (since they carry no current).
Our task therefore is to calculate the current supported by the
states filling the interval $y_{0}^{-} \le y_{0} \le y_{0}^{+}$.
To this end let us first examine the case where the spectrum of filled
states extends over a wider range 
$y_{0}^{--} \le y_{0} \le y_{0}^{++}$ and take 
$\epsilon^{--} \equiv \epsilon(y^{--}_{0}) \gg \epsilon^{-}$
and $\epsilon^{++} \equiv \epsilon(y^{++}_{0}) \gg \epsilon^{+}$
so that the edge states $ y_{0} \lesssim y_{0}^{--}$ and 
$y_{0} \gtrsim  y_{0}^{++}$, traveling extremely fast, 
are well described by the plane-wave states 
of $\hat{H}(\bar{y}_{0},0)$.  
The vacant edge states with $ y_{0} < y_{0}^{--}$ and 
$y_{0} > y_{0}^{++}$ are naturally orthogonal to the occupied
states.

Note that this spectrum of occupied states derives from the
plane-wave states $|y_{0}\rangle$ of $\hat{H}(\bar{y}_{0},0)$ over
the same interval $y_{0}^{--} \le y_{0} \le y_{0}^{++}$ and, 
in view of the completeness of these states, rewrite the sum
$\sum_{\alpha}$ in Eq.~(\ref{Jxnexact}) as a sum over the plane-wave
states $|y_{0}\rangle$, with the result~\cite{fntwo}
\begin{eqnarray} 
J_{x}^{\rm \,wide}&=& -{e\ell^{2}\over{L_{x}}} 
\int^{y_{0}^{++}}_{y_{0}^{--}} dy'_{0}\, \partial_{y'_{0}} 
\langle y'_{0}| \hat{H}(y_{0},k) |y'_{0}\rangle\\
&=& -(e/2\pi)\,(\epsilon^{++} - \epsilon^{--}), 
\label{Jtotal} 
\end{eqnarray}
where the last line is reached by noting that
$\hat{H}(y_{0},k)$ is reduced to the  
plane-wave spectrum $\hat{H}(y_{0},0)$ for $y_{0}\sim y_{0}^{--}$ 
and $y_{0}\sim y_{0}^{++}$; 
we have set $A_{x}=0$ and used the normalization
$\langle y_{0}|y_{0}\rangle=\delta (\ell^{2}p_{x}=0)
=L_{x}/(2\pi \ell^{2})$.

The edge states occupying the intervals 
$y_{0}^{--}\le y_{0}<y_{0}^{-}$ and $y_{0}^{+}< y_{0}\le y_{0}^{++}$
are not necessarily plane-wave states, but the current they carry 
$\propto \int d y_{0}\, \partial_{y_{0}}\epsilon(y_{0})$ is 
exactly calculated from the spectrum:
\begin{equation}
\triangle J_{x}^{\rm \,wide}= -(e/2\pi)\,(\epsilon^{++} - \epsilon^{+} 
+ \epsilon^{-} - \epsilon^{--}).
	\label{trJx}
\end{equation}
Consequently the current carried by the states filling 
the range $y_{0}^{-}\le y_{0}\le y_{0}^{+}$ of our concern turns
out to be determined by the Hall voltage 
$V_{\rm H}= (\epsilon^{+} - \epsilon^{-})/e$ alone:
\begin{equation}
J_{x}^{\{n\}} = -(e^{2}/2\pi)\, V_{\rm H}.
\label{JVH}
\end{equation}
This shows that each filled subband gives rise to 
a single unit $-e^{2}/2\pi\hbar =-e^{2}/h$ of the Hall conductance; 
the emerging Hall plateaus become visible when a significant portion
of electrons gets localized in the sample interior.

Equation~(\ref{JVH}), when combined with the immobility of the
localized states, offers a general but rather indirect proof of
the current compensation theorem.
In the next section we present a more direct proof, based on a
linear-response treatment, which demonstrates an interplay of the
localized states and extended states explicitly. \\

\section{Linear Response }

In this section we examine what would happen to the subband 
described by $[H^{\rm sb}(\bar{y}_{0},k)]_{n n}$ 
when the Hall potential $A_{0}(y)$ is varied by a
small amount $\delta A_{0}(y)$.
The Hamiltonian we consider now is
\begin{eqnarray}
&&H^{\rm new}(\bar{y}_{0},k) = H^{\rm sb}(\bar{y}_{0},k) +
V(\bar{y}_{0},k), \label{Hnew} \\
&&V(\bar{y}_{0},k) =
- e W(\bar{y}_{0},k)\, \delta A_{0}(\bar{y}_{0}\!+\!\ell Y)\, 
W(\bar{y}_{0},k)^{-1}.
\label{WAW}
\end{eqnarray}
The $H^{\rm sb}$ is diagonal in $(m,n)$. 
The potential variation $[V(\bar{y}_{0},k)]_{mn} = 
\delta_{m n}\, \delta A_{0}(\bar{y}_{0}) + \cdots$
in general has off-diagonal pieces which cause transitions to
different subbands. Fortunately, it turns out unnecessary to take
explicit account of such level mixing, as argued below.

The current flowing in response to $\delta A_{0}$ is calculated
from the energy shift to $O(V)$ of each eigenstate $|\alpha \rangle$
of $[H^{\rm sb}]_{nn}$.  The $[H^{\rm sb}]_{nn}$ in general has
degenerate eigenvalues as well as nondegenerate ones.
For a state $|\alpha \rangle$ with a nondegenerate eigenvalue 
the $O(V)$ energy shift is given by 
$\langle\alpha|V(\bar{y}_{0},k)|\alpha\rangle$.
For a degenerate eigenvalue one has to treat all the eigenstates 
belonging to it on an equal footing.
Let $D=\{ |\alpha \rangle \}$ denote one such set of degenerate
states.  The states in $D$ get mixed by the perturbation $V$, and
one can always form, by solving a secular equation, a new
basis $D=\{ |\tilde{\alpha}\rangle \}$ such that $V$ becomes
diagonal in it.  This change of bases 
$D=\{ |\alpha\rangle \} \rightarrow D=\{ |\tilde{\alpha}\rangle \}$
is a unitary transformation and, when $D$ represents inter-level
degeneracy, causes inter-level mixing. 
As is familiar from degenerate perturbation theory, such
$|\tilde{\alpha})$ are a correct zeroth-order choice of basis vectors,
and the $O(V)$ energy shifts of the states 
$|\tilde{\alpha}\rangle$ are given by
$\langle \tilde{\alpha}|V(\bar{y}_{0},k)|\tilde{\alpha}\rangle$.

It is practically impossible to isolate the current component carried
by each of the modes $\tilde{\alpha}$ in $D$ that are still almost
degenerate in energy. It only makes sense to treat each degeneracy set
$D$ as a whole. The effect of the mixing 
$\{ |\alpha\rangle \} \rightarrow \{ |\tilde{\alpha}\rangle \}$
then apparently disappears from the sum of the energy shifts:
\begin{equation}
\sum_{\tilde{\alpha}\in D} \langle \tilde{\alpha}|V(\bar{y}_{0},k)
|\tilde{\alpha}\rangle
= \sum_{\alpha\in D} \langle \alpha|V(\bar{y}_{0},k)|\alpha\rangle .
 \label{}
\end{equation}
This implies that no explicit account of level mixing is needed 
in calculating the Hall current; $V$ does cause inter-subband mixing
but the total Hall current, or the Hall conductance,
is insensitive to it.

In view of this effective independence of each subband, 
we shall henceforth concentrate on a single subband, the $n$th one,
and write $\hat{H}^{\rm new}(\bar{y}_{0},k) \equiv 
[H^{\rm new}(\bar{y}_{0},k)]_{nn}$ and 
$\hat{V}(\bar{y}_{0},k) \equiv [V(\bar{y}_{0},k)]_{nn}$ for short 
so that
\begin{equation}
\hat{H}^{\rm new}(\bar{y}_{0},k) = \hat{H}(\bar{y}_{0},k) + 
\hat{V}(\bar{y}_{0},k).
\end{equation} 
Let us choose anew the eigenstates $|\alpha\rangle$ (with energy
$\epsilon_{\alpha}$)  of $\hat{H}$ so that 
they evolve into the eigenstates $|\alpha\rangle\!\rangle$ of 
$\hat{H}^{\rm new}$ smoothly as $\hat{V}$ is turned on. 
The current 
${\cal J}_{x}^{\{n\}} \equiv J_{x}^{\{n\}}[A_{0}\!+\!\delta A_{0}]$ 
supported by the subband in the presence of the
potential $A_{0} + \delta A_{0}$ is given by Eq.~(\ref{Jxnexact})
with $\hat{H} \rightarrow \hat{H}^{\rm new}$ and 
$|\alpha\rangle \rightarrow |\alpha\rangle\!\rangle$.
Expressing $|\alpha \rangle\!\rangle$ in terms of $|\alpha\rangle$ 
gives rise to the linear-response expression:~\cite{KStwo}
\begin{eqnarray}
{\cal J}_{x}^{\{n\}} &=& -{e\ell^{2}\over{L_{x}}} \sum_{\alpha}\Bigl[
\langle \alpha|\partial_{y_{0}}\hat{H}^{\rm new} |\alpha\rangle\,
+\,{\sum_{\beta}}'{ \rho_{\alpha\beta}
\over{(\epsilon_{\alpha}-\epsilon_{\beta})^{2} }} \Bigr], \nonumber\\ 
\label{JXN} \\  
\rho_{\alpha\beta}&=& (\epsilon_{\alpha}-\epsilon_{\beta})
\left\{ \langle \alpha|\hat{V}|\beta\rangle 
\langle\beta|\partial_{y_{0}}\hat{H}|\alpha\rangle 
+ (\alpha \leftrightarrow \beta) \right\} \nonumber \\
&=& \langle \alpha| [\hat{H}, \hat{V}] |\beta\rangle
\langle \beta|\partial_{y_{0}}\hat{H}|\alpha\rangle
- (\alpha \leftrightarrow \beta),
 \label{rhoab} 
\end{eqnarray}
where the sum $\sum_{\alpha}$ runs over all the occupied states 
$|\alpha\rangle$ and $\sum_{\beta}'$ over all possible
intermediate states $|\beta\rangle$ not degenerate in energy with
$|\alpha\rangle$ within the same subband.  
The $\rho_{\alpha\beta}$, which derive from electron scattering 
by the potential $\delta A_{0}$, are antisymmetric in $(\alpha,\beta)$.

Equation~(\ref{JXN}) is a generalization of the Kubo
formula~\cite{AA} so as to include electron 
edge states and a general potential $A_{0}\!+\!\delta A_{0}$.   
The original formula is recovered 
for electron bulk states in a strong magnetic field 
($W\rightarrow 1$) 
and a uniform field $E_{y}$; in this case, 
$[\hat{H}, \hat{V}]\!\rightarrow 
ie\ell^{2}E_{y}\, \partial_{x}[\hat{U}]_{nn}$,
$\partial_{y_{0}}\hat{H}\rightarrow \partial_{y}\, [\hat{U}]_{nn}$, 
and $\partial_{y_{0}}\hat{H}^{\rm new} \rightarrow eE_{y}$.

Let us again explore consequences drawn by translating the localized
states, with the Hamiltonian 
$\hat{H}^{\rm new}(\bar{y}_{0},k)=\hat{H}^{(0)}(y_{0},k)+\hat{F}$ 
this time, where
\begin{equation}
\hat{F}(y_{0},k) \equiv \hat{V}(\bar{y}_{0},k)
+ e\ell^{2} A_{x} \partial_{y_{0}}\hat{H}^{(0)}(y_{0},k) 
+ \cdots.
\end{equation}  
As before, some care is needed to handle degeneracy inherent to
$\hat{H}^{(0)}=\hat{H}(y_{0},k)$.
We choose its eigenstates $\{ |\alpha) \}$ so that 
$(\alpha|\hat{F}|\beta)$ becomes a diagonal matrix within each set of
degenerate states belonging to the same eigenvalue.
Each state $|\alpha )$ thereby evolves into the corresponding
eigenstate $|\alpha\rangle\!\rangle$ of $\hat{H}^{\rm new}$  
as $A_{x}$ and $\delta A_{0}$ are turned on.
We suppose that during the evolution $|\alpha ) \rightarrow
|\alpha\rangle \rightarrow |\alpha\rangle\!\rangle$ 
each mode $\alpha$ remains either localized or extended.

We go through the procedure of Sec.~IV again: 
Expand the field operators $b_{n}(y_{0},t)$ in terms of the
eigenfunctions of $\hat{H}^{(0)}$, and translate only the localized
modes $b_{\lambda}$, as in Eq.~(\ref{Tb}). 
Equations~(\ref{Hn}) through~(\ref{thttwo}) remain intact, except
for obvious replacement $\hat{H} \rightarrow \hat{H}^{\rm new}$.
This time, however, ${\cal C}_{\lambda\lambda'}$ in
Eq.~(\ref{Olambda}) yields an $O(A_{x})$ energy shift of the form 
for state $\lambda$:
\begin{eqnarray}
{\cal C}_{\lambda\lambda} = ie\ell^{2}A_{x} 
{\sum_{\sigma}}' 
\left[ k_{\lambda\sigma}  \hat{V}_{\sigma\lambda} -
 \hat{V}_{\lambda \sigma} k_{\sigma\lambda} \right] + O(A_{x}^{2}) ,
\label{Cll}
\end{eqnarray}
where $k_{\lambda\sigma}\equiv  (\lambda|k|\sigma)$ and 
$ \hat{V}_{\lambda\sigma}\equiv  (\lambda|\hat{V}|\sigma)$.
The sum $\sum'_{\sigma}$ is taken over all possible extended states 
$|\sigma)$ not degenerate with $|\lambda)$; this is because 
$(\hat{H}^{\rm new})_{\sigma\lambda}=0$ for states $|\sigma)$ and 
$|\lambda)$ belonging to the same eigenvalue.

At the same time, the last two terms in ${\cal H}_{n}$ of
Eq.~(\ref{Hn}) combine to cause 
the $\lambda\rightarrow \sigma\rightarrow \lambda$ virtual 
transitions that give rise to an $O(A_{x})$ energy shift of the form:
\begin{eqnarray}
\triangle \epsilon_{\lambda} &\equiv&{\sum_{\sigma}}' 
(\hat{H}^{\rm new})_{\lambda\sigma} 
(\hat{H}^{\rm new})_{\sigma\lambda}/ 
(\epsilon_{\lambda}-\epsilon_{\sigma}) \nonumber\\
&=& - ie\ell^{2} A_{x} {\sum_{\sigma}}' 
\left[ k_{\lambda \sigma} \hat{V}_{\sigma\lambda} 
-  \hat{V}_{\lambda \sigma} k_{\sigma\lambda} \right] + \cdots, 
\label{dHll}
\end{eqnarray}
where use has been made of the relation
$(\hat{H}^{\rm new})_{\sigma\lambda}= \hat{V}_{\sigma\lambda} 
+ i e\ell^{2}A_{x} k_{\sigma\lambda} 
(\epsilon_{\lambda}-\epsilon_{\sigma}) + O(A_{x}^{2})$.
This $\triangle \epsilon_{\lambda}$ cancels the $O(A_{x})$
terms in ${\cal C}_{\lambda\lambda}$ of Eq.~(\ref{Cll}).  
This proves again that the localized states support no current.

For an extended state $\sigma$ 
the $b_{\sigma}^{\dag} (\hat{H}^{\rm new})_{\sigma\sigma'}
b_{\sigma'}$ term in ${\cal H}_{n}$ of Eq.~(\ref{Hn}) gives rise to  
an $O(A_{x})$ energy shift of the form
$(\partial_{A} \hat{H}^{\rm new})_{\sigma\sigma}$, 
which is to be combined with another $O(A_{x})$ energy shift coming 
from the $\sigma\rightarrow \{\beta\} \rightarrow \sigma$ virtual
transitions. 
Consequently the current ${\cal J}_{x}^{\{n\}}$ is written as  
\begin{equation}
{\cal J}_{x}^{\{n\}} = -{e\ell^{2}\over{L_{x}}} \sum_{\sigma}\Bigl[
(\sigma | \partial_{y_{0}}\hat{H}^{\rm new} |\sigma) 
\,+\,{\sum_{\beta}}'{ \rho_{\sigma\beta}
\over{(\epsilon_{\sigma}-\epsilon_{\beta})^{2} }} \Bigr],  
\label{JXNs} 
\end{equation}
where $\rho_{\sigma\beta}$ is given by the same expression as 
Eq.~(\ref{rhoab}) with $A_{x}\rightarrow 0$. 
[One is free to set $A_{x}\rightarrow 0$ in Eq.~(\ref{rhoab}), 
in which case $\hat{H}\rightarrow \hat{H}^{(0)}$ and 
$|\alpha\rangle \rightarrow |\alpha)$.]
Here the sum $\sum_{\sigma}$ runs over all the occupied extended
states $\sigma$.

The two formulas (\ref{JXN}) and (\ref{JXNs}) are essentially 
the same, except that the immobility of localized states 
is already built in the latter.
Actually, with the characterization of localized states now at 
hand, we can verify their equivalence directly: 
Simply substitute the relation 
\begin{equation}
\langle \beta |\partial_{y_{0}} \hat{H}|\lambda\rangle =
i(\epsilon_{\lambda}-\epsilon_{\beta}) 
\langle \beta |k|\lambda\rangle, 
\label{elemdH}
\end{equation}
valid for a localized state $\lambda$, into Eq.~(\ref{JXN}) 
and rewrite the contribution of the 
$\lambda \rightarrow \{\beta \} \rightarrow \lambda$ transitions as
\begin{eqnarray}
{\sum_{\beta}}'{ \rho_{\lambda\beta}
\over{(\epsilon_{\lambda}-\epsilon_{\beta})^{2} }} 
=-i \langle \lambda|[k, \hat{V}]|\lambda\rangle,
\label{sumrlb}
\end{eqnarray}
which indeed cancels the 
$\langle \lambda| \partial_{y_{0}}\hat{H}^{\rm new}|\lambda\rangle =
i\langle \lambda| [k,\hat{V}]|\lambda\rangle$ term in Eq.~(\ref{JXN}).
A remark regarding this proof is in order here: The immobility of the
localized state $|\lambda\rangle$ can be concluded from 
$i\langle \lambda| [k,\hat{H}]|\lambda\rangle=0$  
or Eq.~(\ref{elemdH}) directly. In our approach, however, 
the vanishing of such expectation values is derived from the response
of the localized states to magnetic translation.  This situation is
quite similar to a statement of Noether's theorem that the existence
of conserved quantities is directly read from the invariance of the
Lagrangian without explicit use of the equations of motion. A symmetry
principle underlying our approach will be clarified in Sec.~VI.

In view of the antisymmetry
$\rho_{\sigma\lambda}= -\rho_{\lambda\sigma}$,  
the $\sigma \rightarrow \lambda \rightarrow \sigma$
and $\lambda \rightarrow \sigma \rightarrow \lambda$ virtual
transitions contribute to ${\cal J}_{x}^{\{n\}}$ in equal magnitude 
but in opposite sign; 
physically this is a manifestation of Fermi statistics.~\cite{KS}
To reveal their interplay
let us again consider the current carried by the states filling 
the interval $y_{0}^{--} \le y_{0} \le y_{0}^{++}$.
One may now take the sums $\sum_{\sigma}$ and
$\sum_{\beta}$ in Eq.~(\ref{JXNs}) simply over these states 
(since the vacant states are plane-wave states orthogonal to them).
Let us  write $\sum_{\beta}=\sum_{\sigma'} + \sum_{\lambda}$.
Then the effects of $\sigma \rightarrow \sigma'\rightarrow \sigma$
transitions combine to vanish. 
On the other hand, the $\sigma \rightarrow \{\lambda\}
\rightarrow \sigma$ transitions,
when summed over all possible extended states $\sigma$, yield
\begin{eqnarray}
\sum_{\sigma} \left(\sum_{\lambda}{ \rho_{\sigma\lambda}
\over{(\epsilon_{\sigma}-\epsilon_{\lambda})^{2} }} \right) 
&=& \sum_{\lambda} \langle\lambda| \partial_{y_{0}} \hat{H}^{\rm new}
|\lambda\rangle.
\label{sumrho}
\end{eqnarray}
which demonstrates how the extended states combine to recover the loss
due to the localized states. 
The current ${\cal J}_{x}^{\{n\}}$ of our concern is now given by the
drift term $\propto \sum_{\alpha} 
(\alpha|\partial_{y_{0}}\hat{H}^{\rm new}|\alpha)$, thus leading to
essentially the same conclusion as Eq.~(\ref{Jtotal}).
Note finally that Eq.~(\ref{sumrho}) can also be derived 
from Eq.~(\ref{sumrlb}); thus the immobility of localized states and
current compensation are correlated.

For completeness we remark that Eq.~(\ref{JXNs}) can also be derived
through a current operator directly; see Appendix C for details.

\section{ Concluding remarks}

In this paper we have examined some general 
consequences of localization in the QHE.
In particular, with a simple characterization of localized electron
states as those with a definite notion of c.m.~position, 
we have shown that (i) the immobility of the localized states and
(ii) exact current compensation by the surviving extended states
follow from the response of the Hall system to
magnetic translation of the localized states alone. 
Our consideration relies heavily on a field-theoretic description 
of Hall electrons in $(n,y_{0})$ space rather than in real $(x,y)$
space. We have noted that the electron edge states and
bulk states are best distinguished in $y_{0}$ space, especially in
terms of the spread of the wave functions in $y_{0}$ space.

It will be important to clarify the symmetry principle underlying our
use of magnetic translation $T=e^{-ie\ell^{2}A_{x}k}$ in 
Secs.~IV and V.  A clue is to note that a special gauge transformation
$\Psi^{\Omega}= \Omega \Psi$ with $\Omega=e^{ie A_{x} x}$ acts 
like $T$. Indeed, $\Omega$ formally removes $A_{x}$ from the
Hamiltonian ${\sf H}$ in Eq.~(\ref{HAU});
${\sf H}^{\Omega}=\Omega {\sf H} \Omega^{-1}$ equals ${\sf H}$ with 
$A_{x} \rightarrow 0$.
Like $T$, this gauge transformation affects extended electron states
substantially [unless~\cite{L} $A_{x} = (2\pi/L_{x}) \times$integers],
and it is clear by now that one cannot conclude from the $A_{x}$
independence of ${\sf H}^{\Omega}$ alone that there is no current.
To reveal the relation between $T$ and $\Omega$ 
let us consider how this $\Omega$ acts on the electron operators
$b_{n}(y_{0},t)$, which are projections of $\Psi$ to each Landau
subband.  Recall that $b= W\,a= W\,G\,\Psi$ in matrix form, 
where the unitary transformation 
$G_{N {\bf x}}=\langle N|{\bf x}\rangle$ takes $\Psi(x,y,t)$
to $a_{n}(y_{0},t)$ and $W(\hat{y}_{0},k)$ projects
$a_{m}(y_{0},t)$ to each exact subband.
Analogously, $\Psi^{\Omega}$ is projected to each subband
so that $b^{\Omega}= W_{0}\,G_{0}\Psi^{\Omega}$, where 
$W_{0}\,G_{0}$ stands for $W\,G$ with $A_{x}\rightarrow 0$.
Therefore the desired gauge transformation law is 
\begin{eqnarray}
 b^{\Omega} &=& W_{0}\, G_{0}\,\Omega\, G^{-1} W^{-1}\, b, \nonumber\\
 &=& W_{0}\, G_{0} G^{-1} e^{ie\ell\, A_{x}(-\ell k + Q)} W^{-1}\, b.
\end{eqnarray}
Noting that $G_{0} G^{-1}= 1  - ie\ell^{2}A_{x} Q +O(A_{x}^{2})$, 
which follows from Eq.~(\ref{dGG}), and that 
$e^{-ie\ell^{2}A_{x} k}  W^{-1} e^{ie\ell^{2}A_{x} k} =
W_{0}^{-1}$, one finds
\begin{eqnarray}
 b^{\Omega} &=& e^{-ie\ell^{2}A_{x} k}\, b  + O(A_{x}^{2}).
\label{bomega}
\end{eqnarray}
(The $a_{n}(y_{0},t)$ also obey the same transformation law.)
This shows that the magnetic translation $T=e^{-ie\ell^{2}A_{x}k}$
is nothing but the gauge transformation $\Omega=e^{ie A_{x} x}$
projected to each Landau subband.
This connection holds to $O(A_{x})$, i.e., to the order relevant to 
our consideration. 
Consequently the immobility of localized states and current
compensation can be formulated upon the assumption that localized
electron states remain essentially unchanged under the gauge
transformation $\Omega$; they are thus consequences of gauge
invariance. 
Actually the magnetic translation $T$ is regarded as a transformation
of a still larger ($W_{\infty}$) gauge symmetry.~\cite{Sak,KS}

Laughlin~\cite{L} gave a simple and general explanation for the QHE
on the basis of localization and gauge invariance. 
Aoki and Ando~\cite{AA}, and Prange~\cite{Pr} also noted the
importance of localization and identified current compensation 
as a key mechanism for the QHE.   
In a sense our approach unites these two approaches and promotes
current compensation to a general phenomenon, resting on the principle
of gauge invariance and valid in the presence of level mixing and edge
states.

The Hall potential $A_{0}(y)$ has so far been treated as 
an external potential, not necessarily weak. 
However, it could equally be a potential induced
internally~\cite{MRB,HT,CSG,T} as a result of charge redistribution
caused by an injected current.  
In reality the current distribution and potential
distribution~\cite{FKHBW} across a sample are determined
self-consistently~\cite{MRB,HT,CSG,T} and $A_{0}(y)$ can be regarded
as such a self-consistent potential.  
In this sense, while no explicit account of the Coulomb interaction
has so far been made, our consideration actually takes it into account
within the Hartree approximation.  
[Remember that our consideration refers to no
explicit form of $A_{0}$, which could even be a general potential 
depending on both $x$ and $y$.]

The current compensation theorem has an implication~\cite{KStwo} on
current distributions in Hall bars. 
It is generally considered~\cite{AA,Pr,L} that electron states remain
extended at the center of each bulk-state spectrum; the current these
states carry flows through the sample bulk. 
The presence of sample edges offers one more possibility: 
It is clear intuitively that, of the electron bulk states, 
those residing near the ``edges of the bulk'' have a better chance of
staying extended than those far in the sample interior;
the key factor here is the difference in topology between the edge and
interior. This suggests that a considerable portion of the current
would flow along the ``bulk edges''. 
The edge current here is a Hall current expelled from the
localization-dominated bulk rather than the edge current carried by
the fast-traveling edge states. 
These two kinds of edge current differ in velocity, direction of flow, 
and channel width.  A numerical experiment is now under way to verify
such an idea of the bulk-edge Hall current; details will be reported
elsewhere.

\acknowledgments

The author wishes to thank Y. Nagaoka and B. Sakita for
useful discussions. 
This work is supported in part by a Grant-in-Aid for 
Scientific Research from the Ministry of Education of Japan, 
Science and Culture (No. 07640398).

\newpage

\appendix

\section{Matrix elements}

In this appendix we derive some formulas relating the matrix 
elements $Y_{mn}=\langle y-\bar{y}_{0} \rangle_{mn}/\ell, 
Q_{mn}=- \ell\, \langle k \rangle_{mn}$, etc., 
to the spectra $\nu_{n}(\bar{y}_{0})$.
Let us recall that $\phi_{n}(y;\bar{y}_{0})$ are the eigenfunctions
of the Hamiltonian $H_{0}=(\omega/2) \bigl[ 
\ell^{2}p_{y}^{2} +(1/\ell^{2})(y - \bar{y}_{0})^{2}\bigr]$
with eigenvalues $\omega(\nu_{n} +1/2)$ and the boundary
condition $\phi_{n}(y=\pm L_{y}/2)=0$.
Consider first the commutator 
$[k, H_{0}]=i(\omega/\ell^{2}) (y-\bar{y}_{0})$ and evaluate its
matrix elements, with the result
\begin{eqnarray}
Y_{n n}&=&-\ell\, {\nu\,}'_{n}(\bar{y}_{0}), \label{ynnA} \\
Q_{m n}&=&i\, Y_{mn}/(\nu_{m}-\nu_{n})\ \ \ \ (m\not=n).
  \label{Qmn}
\end{eqnarray}

Similarly, evaluating the matrix elements of the commutator
$[p_{y},H_{0}]=-i(\omega/\ell^{2})(y-\bar{y}_{0})$ 
with particular attention to integrations by parts yields
\begin{equation}
i(\nu_{m}-\nu_{n})\langle p_{y} \rangle_{m n}+(1/\ell)\, Y_{m n}
=G^{-}_{m n}- G^{+}_{m n},    \label{pyHA}	 
\end{equation}
where $G^{\pm}_{m n}$ refer to the gradients of the wave functions
$\phi_{n}(y;y_{0})$ at the boundary $y=\pm L_{y}/2$:
\begin{equation} 
G^{\pm}_{m n}(\bar{y}_{0})\equiv (\ell^{2}/2)
[(\partial_{y}\phi_{m})(\partial_{y}\phi_{n})]_{y=\pm L_{y}/2}.
\label{GmnA}
\end{equation}
In view of Eqs.~(\ref{ynnA}) and (\ref{pyHA}), these gradients are
related to ${\nu\,}'_{n}$ through $G^{\pm}_{n n} \ge 0$:
\begin{equation}
{\nu\,}'_{n}(\bar{y}_{0})= G^{+}_{n n}(\bar{y}_{0})
- G^{-}_{n n}(\bar{y}_{0}),  \label{nuG}
\end{equation}
which shows that 
${\nu\,}'_{n} > 0$ for $\bar{y}_{0} \sim  + L_{y}/2$ while 
${\nu\,}'_{n} < 0$ for $\bar{y}_{0} \sim - L_{y}/2$.

Finally the commutator $[y-\bar{y}_{0},H_{0}]$ leads to 
\begin{equation}
\langle p_{y} \rangle_{m n}=(i/\ell)(\nu_{m}-\nu_{n}) Y_{m n},
  \label{pyA}
\end{equation}
which is combined with Eq.~(\ref{pyHA}) to give the expression
\begin{equation}
 Y_{m n}=\sigma_{m n}\, \ell\, ({\nu\,}'_{m}{\nu\,}'_{n})^{1/2}
 /[(\nu_{m}-\nu_{n})^{2}-1]  \label{ymnA}
\end{equation}
for $y_{0} \sim \pm L_{y}/2$ 
(where $\nu_{m}-\nu_{n}\not=\pm 1$ in general).
Now $Q_{mn}$ and $\langle p_{y} \rangle_{m n}$ also are known
explicitly. Here the overall sign 
$\sigma_{m n}\equiv {\rm sgn}\{\partial_{y}\phi_{m}\}\,
{\rm sgn}\{\partial_{y}\phi_{n}\}\,{\rm sgn}\{y_{0}\}$ refers to the
signs of $\partial_{y}\phi_{n}$ at the relevant boundary
$y= \pm L_{y}/2$ and $y_{0}\sim \pm L_{y}/2$.  
The standard choice of the bulk-state wave functions leading to
Eq.~(\ref{yplusip}) gives $\sigma_{m n}=1$ for $y_{0} \sim +L_{y}/2$
and $\sigma_{m n}=(-1)^{1+n+m}$ for $y_{0} \sim -L_{y}/2$.
Actually $\sigma_{m n}$ maintains these values 
because it is a topological invariant in each separate edge region
$y_{0}\sim \pm L_{y}/2$, where $\partial_{y}\phi_{n}(y=\pm L_{y}/2)$
are nonvanishing, as seen from Eq.~(\ref{nuG}).

\section{subband Hamiltonians}

In this appendix we outline the derivation of the subband Hamiltonian 
$H^{\rm sb}(\bar{y}_{0},k)$ in Eq.~(\ref{Hsb}).
The Hamiltonian $H(y_{0},k)=\omega (\nu + 1/2) + \bar{U} - e A_{0}$ in
Eq.~(\ref{Hmn}) acts on a whole set of Landau levels spanned by the
bases $\{|n, y_{0}\rangle;  n=0,1,2,\cdots, -\infty <y_{0}< \infty\}$,
and is brought to a diagonal form by a suitable unitary transformation
acting on this whole space.  Inter-level mixing among degenerate
states is thereby properly taken care of. 

Suppose we have determined all the eigenstates of $H(\bar{y}_{0},k)$
via such diagonalization.  Let $\{|\alpha)\}$ denote the whole set of
eigenstates forming the $n$th subband.  
The associated multi-component eigenfunctions 
$\langle m, y_{0}|\alpha)$ (with $m=0,1,2,\cdots$) are functions of 
$\bar{y}_{0}= y_{0} + e\ell^{2} A_{x}$.
We assume that they form a complete set in $y_{0}$ space 
(spanned by a complete $y_{0}$-diagonal basis   
$\{ |y_{0}\rangle; -\infty <y_{0} < \infty\}$);
that is, as in the impurity-free case, 
each subband is taken to be complete with respect to c.m.~motion of 
an electron.

From these (known) eigenfunctions one can construct 
$H^{\rm sb}(\bar{y}_{0},k)$ in the following way:
In the $\{ |\alpha) \}$ basis the coordinate operator $y_{0}$ is 
represented by a matrix $(\alpha |y_{0} | \alpha')$, which may be 
diagonalized to give a complete set of wave functions 
$\langle y_{0}|\alpha ) \equiv f_{\alpha}(\bar{y}_{0})$ 
in the $\{ |y_{0}\rangle \}$ basis.  
In this $\{ |y_{0}\rangle \}$ basis the Hamiltonian for the $n$th
subband is expressed, in terms of eigenvalues  $\epsilon_{\alpha}$,
as a matrix:
\begin{equation}
\langle y'_{0}| [H^{\rm sb}]_{nn}|y''_{0}\rangle 
=\sum_{\alpha}  \epsilon_{\alpha} f_{\alpha}^{*}(\bar{y}'_{0}) 
f_{\alpha}(\bar{y}''_{0}),
\end{equation}
which is a function of $\bar{y}'_{0}= y'_{0} + e\ell^{2} A_{x}$ and 
$\bar{y}''_{0}= y''_{0} + e\ell^{2} A_{x}$.
This matrix is readily cast in an operator form.
Indeed, with an expansion of the form 
$\sum_{\alpha}  \epsilon_{\alpha} f_{\alpha}^{*}(\bar{y}'_{0}) 
f_{\alpha}(\bar{y}''_{0})
=\sum_{i,j} c_{ij}\, \phi_{i}(\bar{y}'_{0})\, \phi_{j}(\bar{y}''_{0})$
in terms of the eigenfunctions $\phi_{i}(\bar{y}_{0})$
$(i=0,1,2,\cdots$) of the harmonic oscillator formed of 
$d=(\bar{y}_{0}/\ell + i\ell\, k)/\sqrt{2}$ and 
$d^{\dag}$ with $[d, d^{\dag}]=1$, one obtains the expression
\begin{equation}
	[H^{\rm sb}(\bar{y}_{0},k)]_{nn} = 
        \sum_{i,j=0}^{\infty}  (c_{ij}/\sqrt{i!j!})\,
	: (d^{\dag})^{i} d^{j} e^{ -d^{\dag}d}: \ ,
\end{equation}
where normal ordering such that $:d d^{\dag}:\, = d^{\dag} d$ is 
understood.  (For a check it is enlightening to verify that the choice
$c_{ij}= i\,\delta_{ij}$ recovers the number operator $d^{\dag}d$.)
Likewise one can derive the operator expression for $W(\bar{y_{0}},k)$
from the matrix elements:
\begin{equation}
\langle m,y_{0}| W^{-1}|n,y'_{0} \rangle 
= \sum_{\alpha} \langle m,y_{0}|\alpha)\, f^{*}_{\alpha}(\bar{y}'_{0}).
\end{equation}

\section{Current operators}

We have passed from $a_{m}(y_{0},t)$ to $b_{n}(y_{0},t)$ via a unitary
transformation $[W(\bar{y}_{0},k)]_{nm}$, which, being a finite
functional of the potential $U -eA_{0}$, is a well-defined
operator.  One can therefore adopt
\begin{equation}
{J_{x}}'= - (1/L_{x})\, 
b_{\alpha}^{\dag}\, (\partial_{A}\hat{H})_{\alpha\beta}\, b_{\beta}
	\label{Jxbb}
\end{equation}
as a current operator physically equivalent to $J_{x}$ in
Eq.~(\ref{Jxx}); $(\partial_{A}\hat{H})_{\alpha\beta}
=(\alpha|\partial_{A}\hat{H}|\beta)$.

Let us arrange $\hat{H}_{\lambda\lambda'}, \hat{H}_{\lambda\sigma'},  
\hat{H}_{\sigma\lambda'}, \hat{H}_{\sigma\sigma'}$ 
in the form of a $2\times 2$ matrix $\hat{H}^{\rm mat}$ acting on 
$(b_{\lambda'},b_{\sigma'})^{\rm t}$, 
and also $k_{\lambda\lambda'}$ in an analogous matrix form $K$. 
Then the commutator $[K, \hat{H}^{\rm mat}]$, written in the form
\begin{eqnarray}
\triangle{\cal H}_{n}
&=& b^{\dag}_{\lambda}\, [\, k_{\lambda\lambda''}
\hat{H}_{\lambda''\lambda'} - \hat{H}_{\lambda\lambda''}
k_{\lambda''\lambda'} ]\, b_{\lambda'} \nonumber \\
&&  + b^{\dag}_{\lambda} k_{\lambda\lambda'} \hat{H}_{\lambda'\sigma}
b_{\sigma} - b^{\dag}_{\sigma}  \hat{H}_{\sigma \lambda} 
k_{\lambda\lambda'} b_{\lambda'},
\label{dTH}
\end{eqnarray}
has a vanishing expectation value for each eigenstate of
$\hat{H}$.  One may thus add it to the current ${J_{x}}'$ without 
modifying the physical content.  
Actually the translation~(\ref{Tb}) takes us to the redefined current
$J_{x}^{\rm red}={J_{x}}'+ (e\ell^{2}/L_{x}) \triangle {\cal H}_{n}$.
It is now a simple exercise to verify the current compensation theorem
directly with $J_{x}^{\rm red}$.  
[Set $\hat{H}\rightarrow \hat{H}^{\rm new}$ in $J_{x}^{\rm red}$
to get to Eq.~(\ref{JXNs}).]

\bigskip


\begin{figure}
\epsfxsize=8cm
\epsfysize=5cm
\centerline{\epsfbox{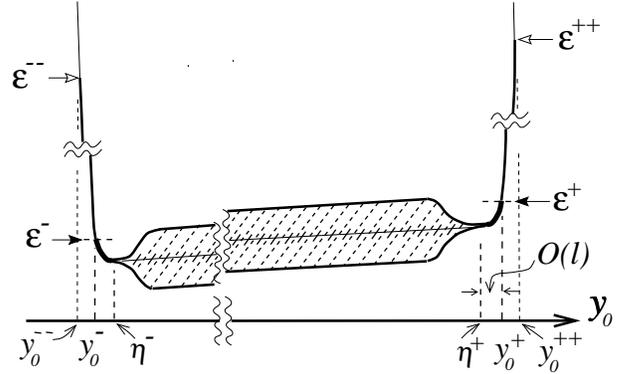}}
\vspace{0.15in}
\caption{ Energy spectrum of an impurity-broadened Landau subband.  
The bulk-state spectrum lies in the interval 
$\eta^{-} \le y_{0}\le\eta^{+}$.
The edge-state spectrum is drawn with solid curves.}
\label{fig1}
\end{figure}



\begin{thebibliography}{99}
\bibitem{V} K. von Klitzing, G. Dorda and M. Pepper, Phys. Rev. Lett.
{\bf 45}, 494 (1980); D. C. Tsui, H. L. Stromer and A. C. Gossard, 
Phys. Rev. Lett. {\bf 48}, 1559 (1982).

\bibitem{Rev} For a review see, {\em The Quantum Hall Effect}, 
edited by R.E. Prange and S.M. Girvin (Springer-Verlag, Berlin, 1987).

\bibitem{AA} H. Aoki and T. Ando, Solid State Commun. {\bf 38}, 1079 (1981).
See also, T. Ando, Y. Matsumoto, and Y. Uemura,
J. Phys. Soc. Jpn. {\bf 39}, 279 (1975).

\bibitem{Pr} R. E. Prange, Phys. Rev. B {\bf 23}, 4802 (1981).
\bibitem{L} R. B. Laughlin, Phys. Rev. B {\bf 23}, 5632 (1981).
\bibitem{H} B. I. Halperin, Phys. Rev. B {\bf 25}, 2185 (1982).
\bibitem{TK} D. J. Thouless, M. Kohmoto, M. N. Nightingale, 
and M. den Nijs, Phys. Rev. Lett. {\bf 49}, 405 (1982);
Q. Niu, D. J. Thouless, and Y.-S. Wu, Phys. Rev. B {\bf 31}, 3372
(1985); Q. Niu and D. J. Thouless, Phys. Rev. B {\bf 35}, 2188 (1987).

\bibitem{LLP} H. Levine, S. B. Libby and A. M.M. Pruisken, 
Nucl. Phys. B {\bf 240} [FS12], 49 (1984).

\bibitem{MS} A. H. MacDonald, and P. Streda,
Phys. Rev. B {\bf 29}, 1616 (1984).

\bibitem{SKM} P. Streda, J. Kucera and A. H. MacDonald,
 Phys. Rev. Lett. {\bf 59}, 1973 (1987);
J. K. Jain and S. A. Kivelson, Phys. Rev. B {\bf 37}, 4276 (1988);
M. B\"uttiker, Phys. Rev. B {\bf 38}, 9375 (1988).

\bibitem{Wen} X.-G. Wen,  Phys. Rev. B {\bf 43}, 11025 (1991);
M. Stone, Ann. Phys. (N.Y.) {\bf 207}, 38 (1991).

\bibitem{KS} K. Shizuya, Phys. Rev. B {\bf 45}, 11143 (1992);
Phys. Rev. B {\bf 52}, 2747 (1995).

\bibitem{KStwo} K. Shizuya, Phys. Rev. Lett. {\bf 73}, 2907 (1994).

\bibitem{WW} E. T. Whittaker and G. N. Watson, 
{\sl A Course of Modern Analysis},
(Cambridge University Press, London, 1927).

\bibitem{fn} This $\hat{U}(\bar{y}_{0},k)$ is equivalent to an
alternative expression $U_{m n}$ used earlier in Ref. \onlinecite{KStwo}.

\bibitem{Zak} J. Zak,  Phys. Rev. {\bf 134}, A1602 (1964).

\bibitem{P} R. E. Peierls, {\em Surprises in Theoretical Physics}, 
(Princeton University Press, Princeton, 1979).

\bibitem{fntwo} Note 
$\langle y'_{0}|\partial_{y_{0}}\hat{H}(\bar{y}_{0},k)|y''_{0}\rangle
 = (\partial_{y'_{0}}+\partial_{y''_{0}}) 
\langle y'_{0}|\hat{H}(\bar{y}_{0},k)|y''_{0}\rangle$.

\bibitem{Sak} B. Sakita, Phys. Lett. B {\bf 315}, 124 (1993).

\bibitem{MRB} A. H. MacDonald, T. M. Rice, and W. F. Brinkman,
 Phys. Rev. B {\bf 28}, 3648 (1983).

\bibitem{HT} O. Heinonen and P. L. Taylor, 
 Phys. Rev. B {\bf 32}, 633 (1985).

\bibitem{CSG} D. B. Chklovskii, B. I. Shklovskii, and L. I. Glazman,
Phys. Rev. B {\bf 46}, 4026 (1992).

\bibitem{T} D. J. Thouless, Phys. Rev. Lett. {\bf 71}, 1879 (1993);
C. Wexler and D. J. Thouless, Phys. Rev. B {\bf 49}, 4815 (1994).

\bibitem{FKHBW} P. F. Fontein, et al., Phys. Rev. B {\bf 43}, 
12090 (1991).\\
\end{thebibliography}
\end{document}